Single crystal growth, structure and electronic properties of metallic delafossite PdRhO$_2$

P. Kushwaha [1], H. Borrmann [1], S. Khim [1], H. Rosner [1], P.J.W. Moll [1], D. A. Sokolov [1], V. Sunko [1,2], Yu. Grin [1] and A.P. Mackenzie [1,2]

[1] *Max Planck Institute for Chemical Physics of Solids, Nöthnitzer Str. 40, 01187 Dresden, Germany.*
[2] *Scottish Universities Physics Alliance, School of Physics and Astronomy, University of St. Andrews, St. Andrews KY16 9SS, U.K.*

**Abstract**

We report growth of single crystals of the non-magnetic metallic delafossite PdRhO$_2$, comparing the results from three different methods. Complete crystallographic data were obtained from single crystal X-ray diffraction, and electronic structure calculations made using the refined structural parameters. Focused-ion beam microstructuring was used to prepare a sample for measurements of the in- and out-of-plane electrical resistivity, and the large observed anisotropy is qualitatively consistent with the cylindrical Fermi surface predicted by the calculations.

**Introduction**

Although they were first synthesized nearly fifty years ago [1–3], the intriguing properties of layered metals with the delafossite structure (Fig. 1) have only been fully appreciated in work done in the past decade [4]. First, non-magnetic $PdCoO_2$ [5,6] and magnetic $PdCrO_2$ [7] were shown to have extremely high conductivities for in-plane (perpendicular to the crystallographic [001] direction) electrical transport, and large anisotropies of factors of several hundred between in-plane and out-of-plane (parallel to [001]) conductivity. Next, relatively large single crystals of non-magnetic $PtCoO_2$ were synthesized [8], and the Fermi surfaces and electronic structures of the three materials established by calculation [9–12] as well as by direct measurements of angle-resolved photoemission spectroscopy [8,13–15] and the de Haas-van Alphen effect [12,16,17]. The picture that has emerged from this research is fascinating. In the non-magnetic materials, conduction is based on a single, extremely broad band of (Pd,Pt) origin and a cylindrical Fermi surface with hexagonal cross-section and very weak corrugation along $k_z$. Consistent with such a broad band, the measured Fermi velocities are close to the free electron value, even though the band clearly has substantial *d* orbital content. The situation is made even more interesting by observations that confirm strong electronic correlations in the (Co,Cr) layers, meaning that the metallic delafossites offer a 'natural heterostructure' (natural intergrowth) of planes of nearly free electrons and strongly correlated insulators.

Quantitative analysis of the conduction properties further highlights why the metallic delafossites merit further study. At room temperature, the resistivities of $PdCoO_2$ and $PtCoO_2$ are 2.6 μΩcm and 2.1 μΩcm respectively [8,16], lower than that of almost all elemental metals. Taking into account the lower carrier concentration of the delafossites, the room temperature mean free paths of nearly $10^3$ Å are longer than those of even Cu or Ag, so the delafossites are conductivity record-holders [4]. At low temperatures, mean free paths as long as 2 x $10^5$ Å have been reported in $PdCoO_2$, an astonishingly large value for a metallic oxide, and one that has opened the possibility of a range of exotic experiments in the high purity limit [18–20].

Although we now know that the metallic delafossites have fascinating properties, it is less clear why this should be so. The broad conduction band appears to be related to *s-d* hybridization of the (Pd,Pt) states, and spin-orbit coupling is important in stabilizing the simple Fermi surface [8]. Electronic structure calculations also indicate that oxygen positions and the strength of correlations in the transition metal layers might play an important role [8], but more work will be required to establish how important these effects are. What is very clear is that the insights that have been obtained so far have required work on high quality single crystals, and that there is a strong motivation to extend the number of materials for which such crystals are available.

Before describing the new crystal growth work that we have performed, it is instructive to review past work on the preparation of metallic delafossites. Rogers *et al*. first patented the synthesis of the Pd based conducting delafossites Pd(Co,Cr,Rh)$O_2$ in 1970 by using different combinations of reaction of noble metal (Pd, Pt), noble metal halides ($PdCl_2$, $PtCl_2$)/oxides (PdO, PtO) along with transition metal oxide precursors (e.g. CoO, $Co_3O_4$) in sealed ampoules [21,22]. They followed this up with detailed reports of the crystal growth of $PdCoO_2$ using the reaction of $PdCl_2$ and CoO at 700 °C in a sealed ampoule [21,22] resulting in a crystal size of 0.5 mm on an edge. $PtCoO_2$ single crystals were grown by hydrothermal reaction of Pt and $Co_3O_4$ with concentrated hydrochloric acid under applied pressure of 3000 atm at 500 - 700 °C [1,22]. This procedure resulted in crystals with dimensions up to approximately one mm,. In 1997, Tanaka *et al.* [23] reported growth procedures for $PdCoO_2$ and $PtCoO_2$ at ambient pressure. They used the same reaction of (Pd,Pt)$Cl_2$ and CoO under vacuum but mentioned that the crystal size of (Pt, Pd)$CoO_2$ was proportional to the starting ratio of (Pt, Pd)$Cl_2$ and CoO. The crystal size achieved was approximately one mm for $PdCoO_2$, but only 40-50 μm for $PtCoO_2$. Recently, using a similar technique but a change in temperature profile (such as maximum temperature for reaction, dwelling temperature/time and cooling rate), a further increase in crystal dimension to approximately 2 mm was achieved both for $PdCoO_2$ [20] and $PtCoO_2$ [8]. Takatsu and Maeno [7] reported that a solid state reaction followed by a salt flux growth technique led to the growth of PdCrO$_2$ crystals with a dimension of 4-5 mm in plane and 0.5 mm along the *c*-axis).

The subject of this paper is the next known metallic compound in the series, PdRhO$_2$. Although it was synthesized in powder and polycrystalline thin film [24,25] forms as part of the Dupont group's seminal research in the 1970s and 1980s, no single crystal growth has, to our knowledge, been reported. Almost nothing is known experimentally about its properties, although one electronic structure calculation [26] has predicted it to be metallic with a Fermi surface similar to those of PdCoO$_2$ and PtCoO$_2$, and another successfully modeled its Raman-active modes [27]. In order to investigate the presumed anisotropic transport properties of PdRhO$_2$, and eventually to study its electronic structure with, for example, angle resolved photoemission spectroscopy, it is vital that single crystals of reasonable size be available. Here, we report the use of several techniques to grow crystals of PdRhO$_2$. We perform a precise single crystal structure refinement and preliminary transport data on a device produced by focused ion beam (FIB) microstructuring, and close by presenting calculations to investigate the importance of structural details in determining the electronic band structure.

**Crystal growth**

**Method 1:** We started with a similar procedure to that reported by Shannon *et al.*[1] using LiRhO$_2$ as shown in equation 1.

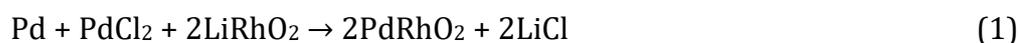
Pd + PdCl$_2$ + 2LiRhO$_2$ → 2PdRhO$_2$ + 2LiCl      (1)

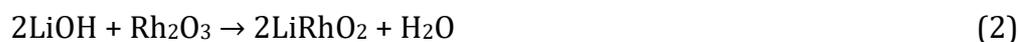
2LiOH + Rh$_2$O$_3$ → 2LiRhO$_2$ + H$_2$O      (2)

However, we found that the final product always contained elemental Rh. In order to prepare single-phase stoichiometric LiRhO$_2$ we used a mixture of lithium hydroxide and rhodium oxide powder (prepared by heating 99.99% pure Rh powder at 950 °C for 24 hours in oxygen flow; the resulting oxide was confirmed to be Rh$_2$O$_3$ by X-ray diffraction) as shown in equation 2 and heated at 850 °C for 48 hours in oxygen flow (to prevent the reduction of rhodium toward Li$_x$RhO$_2$ at elevated temperatures) followed by cooling to room temperature at a rate of 1 °C/min. This preparation method gave us single-phase and stoichiometric

LiRhO$_2$. This stoichiometric LiRhO$_2$ was thoroughly ground with palladium powder (99.95%) and palladium chloride powder (99.999%) inside a glove box under Ar. The ground powder mixture was sealed in a quartz tube under vacuum and heated to 930 °C in 3 hours then taken up to 1000 °C over the next 6 hours, cooled down to 800 °C over two hours and held there for 40 hours followed by further cooling to room temperature in 3 hours. The final product was a cluster of microcrystals. It was washed with aqua regia and distilled water many times to remove by-products. A high-resolution scanning electron microscopy (SEM) image of this washed sample of PdRhO$_2$ microcrystals is shown in Fig 2. This image shows well-shaped hexagonal nanocrystals of PdRhO$_2$, for which the crystal size varies from a few hundred nanometers to couple of microns.

Although too small for most single crystal experiments, the material from this first experiment could readily be powdered for X-ray diffraction. In Fig 3 we show the Rietveld refinement of room temperature X-ray diffraction data taken on finely ground powder (Cu $K\alpha_1$ radiation, $\lambda$ = 1.54056 Å). The fit to model of the delafossite type structure (space group $R\bar{3}m$) is excellent, and the lattice parameters of $a$ = 3.0240(6) Å and $c$ = 18.096(6) Å agree well with those found in the literature, as shown in the Table 1, where we summarize the lattice parameters for the growth following methods 1 and 2.. We attempted to obtain bigger crystals by changing the flux from the PdCl$_2$ mentioned above to NaCl, LiCl and PdBr$_2$, but obtained only fine powder instead of single crystals.

**Method 2:** Since changing the flux was unsuccessful, we decided to make a different selection of the precursor materials and perform the reaction using Rh$_2$O$_3$ instead of LiRhO$_2$, in a process similar in spirit to the previous successful growth of PdCoO$_2$ from PdCl$_2$ and CoO [1,3]. Rh$_2$O$_3$ powder was mixed with Pd and PdCl$_2$ (equation 3) inside the glove box and the mixture was sealed in a quartz tube under vacuum.

Pd + PdCl$_2$ + $x$Rh$_2$O$_3$ → 2PdRhO$_2$ + byproducts          (3)

The tube was then heated in a vertical furnace employing the same temperature profile as used for method 1. Multiple attempts highlighted the importance of parameter *x* in equation 3. If *x* is too small, few PdRhO$_2$ crystals are formed, accompanied by unwanted byproducts. After many trials, we found that *x* = 1.5 gave a reasonable yield of bigger crystals, which were confirmed to be stoichiometric within experimental error by electron probe microanalysis. Fig 4 shows optical and SEM images of as-grown crystals. The average crystal size is about 100 μm in the *ab*-plane. Usually, delafossite crystals grow as platelets whose thickness along the *c*-axis is very small, which limits measurements of the *c*-axis properties. However, the crystals grown here using method 2 look like bulky three dimensional crystals where the thickness along the *c*-axis is comparable to the typical size in the *ab*-plane. Even though these crystals may still not be large enough for some physical measurements, they were adequate for micro fabricated devices which can be used in electrical transport and other measurements.

| growth condition | type of diffraction | ref. | a (A) | c (A) |
|---|---|---|---|---|
| method1 | powder | Shannon et al(1971) | 3.0209(2) | 18.083(2) |
| method1 | powder | this work | 3.0240(6) | 18.096(6) |
| method2 | powder | this work | 3.0235(6) | 18.094(9) |
| method2 | single-crystal | this work | 3.0251(3) | 18.101(3) |

Table1. Comparison of the lattice parameters of PdRhO$_2$ determined by the powder and single crystal X-ray diffraction in this work and Ref.1.

**Method 3:** Further increase in size of the single crystals was realized by adopting a temperature gradient like that used in the chemical vapor transport (CVT) method. The mixture of PdCl$_2$ and Rh$_2$O$_3$ was sealed in an evacuated quartz ampoule ≈ 10 cm long. In a two-zone horizontal tube furnace, the ampoule was heated to 1000 °C for 3 days and each zone were respectively set to $T_{\text{hot}}$ = 700 °C (at the location of the initial mixture) and $T_{\text{cold}}$ = 670 °C. This temperature gradient was maintained for 10 days, after which the furnace was cooled down to room temperature over 20 hours. As shown in Fig. 5(a), single crystals [Fig. 5(b)] with a lateral length of ∼ 0.5 mm were collected out of the reacted mixture from the hot zone, and the unreacted PdCl$_2$ was condensed in the cold zone. The remaining charge such as Rh$_2$O$_3$, PdCl$_2$, and RhCl$_3$ on the crystal's surface was removed in aqua regia acid solution (HCl : HNO$_3$ = 1 : 1 in volume) for 5 min. Care needed to be taken in determining the thermal parameters for the growth. While Rh$_2$O$_3$ tends to decompose into pure Rh for a long dwelling time and oxygen poor atmosphere at an elevated temperature, the formation of PdRhO$_2$ phase requires a certain minimum temperature. Compromising between those competing factors gives rise to a proper temperature window for growth. In addition, owing to the volatile nature of Rh$_2$O$_3$ and PdCl$_2$, we realize a mineralization in a vapor form under a specific temperature gradient. This modification leads to the enhancement of the crystal size.

**Electrical resistivity measurements**

Since the yield of the larger crystals shown in Fig. 5 was fairly low, and the bigger samples were saved for thermodynamic and spectroscopic measurements, in order to perform first electrical resistivity measurements we used focused ion beam (FIB) techniques to prepare the sample shown in Fig. 6. Using techniques similar to those employed for preparing lamellae for transmission electron microscopy, a slice approximately 150 x 50 x 2 μm was cut from a larger crystal, such that the large face was spanned by the crystallographic *a* and *c* axes. This was then laid on insulating epoxy, and cut into the shape shown in Fig. 6. In this device, current flows between contacts I$_1$ and I$_2$, voltage pairs V$_{1,2}$ and V$_{5,6}$ in principle enable a measurement of the *c*-axis conductivity, and V$_{3,4}$ allow a measurement of the conductivity in-plane. The contacts to the external wires are made with FIB-deposited Pt, in contact with FIB sculpted plates of the same starting slice as the conducting track.

The results of the transport measurements are shown in Fig. 7. The in-plane resistivity at 300 K is just under 10 μΩcm, a factor of 4-5 higher than that of PdCoO$_2$ or PtCoO$_2$ and comparable with that seen in PdCrO$_2$ in which magnetic scattering plays a role not expected in PdRhO$_2$. An in-plane resistivity ratio ($\rho_{300K}/\rho_{4K}$) of only about 10 is seen. This is far smaller than that reported in bulk crystals of the other metallic delafossites, for which they are usually in excess of 100. However, we caution that this might well be a large underestimate. The low temperature value may be dominated by boundary scattering because the thickness of the sample (which determines the boundary separation for in-plane transport) is only 1.6 ± 0.2 µm. For tracks of this width, the resistivity ratio of $PdCoO_2$ drops from over 300 to approximately 25, an effect shown to be due to boundary scattering rather than bulk defects or damage from the FIB irradiation [19]. In $PtCoO_2$ unpublished data from our group shows in-plane resistivity ratio dropping from 53 for 8 µm track width to between 10 and 15 for this track width. In this context, the value seen in Fig. 7a) in $PdRhO_2$ almost certainly has a strong contribution from boundary scattering, and would be of order 100 in a bulk crystal. Even though some overlooked systematic errors may affect the detailed values that we quote for the resistivity in the two directions, the experiment clearly demonstrates the qualitative fact that $PdRhO_2$ has a high metallic conductivity in the conducting planes, and a large anisotropy between in- and out-of-plane electrical transport.

**X-ray diffraction and structure refinement**

The existence of high quality single crystals offers the opportunity to perform a structure refinement for $PdRhO_2$ of similar quality to those available for other delafossites, so we set out to do this. A small crystal (40 × 33 × 28 µm$^3$) was chosen after thorough investigation under an optical microscope and used for single crystal X-ray measurements on a Rigaku AFC7 diffractometer with a Saturn 724+ CCD detector. After preliminary unit cell determination, oscillation images around the unit cell axes (shown in Fig. 8) proved very good crystal quality without any indication of partial cleavage or twinning. All diffraction experiments were performed at 295 K applying graphite-monochromated Mo$K\alpha$ radiation ($\lambda$ = 0.71073 Å) collimated with a mono-capillary. A total of five full φ-scans resulted in 3100 images from which after integration and scaling 2469 Bragg intensities were obtained. After averaging, 218 unique reflections were used in structure refinement. Derived lattice parameters $a$ = 3.0251(3) Å and $c$ = 18.101(3) Å are in very good agreement with literature data as well as with our powder data. Refinement of the established model in the space group $R\bar{3}m$ (3Pd in 3(*a*) 0 0 0, $U_{eq}$= 0.00661(8) Å$^2$, 3Rh in 3(*b*) 0 0 ½, $U_{eq}$ = 0.00397(8) Å$^2$; 6O in 6(*c*) 0 0 $z$ = 0.11008(11), $U_{eq}$ = 0.0064(3) Å$^2$) converged in an excellent fit of 9 parameters vs. all 218 independent reflections. Agreement based on $F^2$ including an isotropic extinction correction is indicated by $wR2$= 0.0376 and goodness-of-fit of 1.63. Although Pd and Rh are hardly distinguishable using X-ray methods, tests of the refinement reveal quite surprising results. For the model where both metal sites are occupied by Pd or by Rh only, the fit is very similar at $wR2$ = 0.0449 (Pd) vs. 0.0447 (Rh), however, when both metal positions are interchanged the fit becomes considerably worse ($wR2$ = 0.0616). As well as providing independent verification of the established structural model, we take this as a clear proof for the outstanding quality of the data and thus confirmation of the excellent quality of the crystal under investigation. In comparison to PdCoO$_2$ and PtCoO$_2$ the a-axis is considerably larger for PdRhO$_2$ which is a direct consequence of increased Rh–O distance vs. Co–O distances. Table. 2 gives respective values for four established delafossite metals. Interestingly, PdRhO$_2$ shows close similarities to PdCrO$_2$, in particular with respect to Pd–O distances.

|            | PdCoO$_2$ | PtCoO$_2$ | PdRhO$_2$ | PdCrO$_2$ |
|------------|-----------|-----------|-----------|-----------|
| $B$ – O [Å] | 1.908     | 1.89      | 2.025     | 1.970     |
| $A$ – O [Å] | 1.973     | 2.03      | 1.993     | 1.999     |

Table. 2 Comparison of selected structural details of four delafossite metals with a generic formula ABO$_2$, where A=Pd, Pt, and B=Co, Rh, Cr. The structural details for PdCoO$_2$, PtCoO$_2$, and PdCrO$_2$ were determined from the published crystal structure data[1].

**Electronic structure calculations**

Since the refined structure reported here differs significantly in detail from those previously published, it is important to check for the effect that these changes have on the electronic structure. In order to do this, we performed comparative relativistic density functional (DFT) calculations using the full-potential FPLO code [28,29]. For the exchange-correlation potential, within the general gradient approximation (GGA), the parametrization of Perdew-Burke-Ernzerhof [30] was chosen. To obtain precise band structure information, the calculations were carried out on a well converged mesh of 27000 $k$ points (30 × 30 × 30 mesh, 2496 points in the irreducible wedge of the Brillouin zone). For all calculations, the experimental crystal structure reported above was used.

The resulting band structure near the Fermi level $E_F$ is shown in Fig. 9. In agreement with ref. [26], we find a single band crossing the Fermi level, resulting in a cylindrical Fermi surface. However, although the differences in the structural parameters seem small - only 1% in the lattice parameter and 1 degree in the rhombohedral angle [#] - we find substantial differences of detail in the resulting band structures: (i) Our bandwidth for the band crossing $E_F$ is about 20% larger compared to ref. [26], in consequence yielding (ii) a density of states (DOS) at $E_F$ that is approximately 20% smaller and (iii) a plasma frequency $\omega_P$ that is approximately 10% larger and which exhibits (iv) approximately a 15% larger anisotropy for the in-plane versus out-of-plane components. (v) The total bandwidth of the valence band is approximately 0.5 eV larger than in the calculation of ref. [26]. The band structure calculated for the lattice parameters of ref. [3] is in between the other two. We also studied the influence of the different internal O position parameters for the different structure data sets and found it to be of minor relevance; the changes in electronic structure are dominated by the measured differences in lattice parameter. As stated in ref. [26], we find a small influence of the spin orbit coupling (SOC) near $E_F$, but band splittings of the order of 100 meV for the lower lying bands.

The calculated electronic DOS is shown in Fig. 10. In general, the DOS of the valence band is very similar to those of the closely related compounds $PdCoO_2$ and $PtCoO_2$: The bandwidth of $PdRhO_2$ and $PdCoO_2$ is essentially the same and about 1 eV narrower than of $PtCoO_2$. However, the Rh dominated states in $PdRhO_2$ (between - 0.5 and - 2.5 eV) are significantly broader (about 0.6 eV) then the respective Co states in the other compounds. This originates from the less localized nature of the Rh 4*d* orbitals compared to Co 3*d*, resulting in a stronger Rh-O hybridization. This stronger hybridization is also responsible for the gap in the DOS above 0.5 eV in contrast to the Co-systems.

---

[#] To ease a comparison with the structural parameters used in the two calculations, our lattice parameters in a rhombohedral setting are $a$ = 6.2814 Å, $\alpha$ = 27.87° compared to $a$ = 6.32 Å, $\alpha$ = 28.12° quoted in ref. [26].

Also with respect to the Fermi surface, all three systems are rather similar, exhibiting a quasi two-dimensional Fermi surface originating from a single, half-filled band. This band is dominated by Pd-4*d* states. However, due to the smaller contribution of Rh (compared to Co) the shape of the band is slightly different, in particular at the band top, as can be seen by comparing Fig. 9 of the current manuscript with Fig. 2 in ref. [26] and Fig.4 in ref. [8].

We conclude from our calculation that precise structural data are crucial to understand the details of the electronic structure. This will be of importance for future electronic structure studies, in particular with respect to the separation of many body interactions from the predictions based on "single particle bands".

**Conclusions**

In summary, we have employed several experimental techniques for the growth of single crystals of a material, $PdRhO_2$, which is of considerable topical interest. To our knowledge these are the first successful growths of crystals of this material, and our work points to mineralization of gas phases being a promising method for the growth of larger crystals. Our structure determination, electrical resistivity measurements and electronic structure calculations establish $PdRhO_2$ to be the third known metallic delafossite whose low temperature Fermi surface is a single cylinder. As such, we hope that it will be an ideal material on which to develop and benchmark our understanding of this new and rapidly growing field.

**Figure captions**

Fig. 1. The crystal structure of PdRhO$_2$, highlighting the RhO$_6$ octahedra and the unusual linear O-Pd-O groups perpendicular to the Pd layers.

Fig. 2. A high resolution scanning electron microscope image showing the clearly faceted PdRhO$_2$ microcrystals prepared with method 1 (see main text).

Fig. 3. Room temperature X-ray diffraction pattern of PdRhO$_2$ powder prepared by methods 1-3 (Cu-K$\alpha_1$ radiation, $\lambda$ = 1.54056 Å, see main text for details). The symbols show the observed intensities fitted using the Rietveld refinement (black line). Vertical lines show Bragg peak positions and the blue line at the bottom corresponds to the difference between the data and the fit. Powder x-ray diffraction patterns obtained from samples grown by the methods 2 and 3 were essentially identical, except for the presence of small fraction of unreacted Rh$_2$O$_3$ in the sample prepared by the method 2.

Fig 4 (a) Optical image, (b) and (c) SEM images of PdRhO$_2$ crystals grown by method 2 as described in the main text.

Fig. 5 (a) The quartz ampoule used for method 3, as described in the main text, after heating. The grown single crystals were found in the reacted mixture at the hot zone and byproducts, mainly PdCl$_2$, were found at the cold zone. (b) The collected single crystals have the characteristic hexagonal shape commonly observed for delafossites.

Fig. 6. The micro-structured PdRhO$_2$ plate described in the main text.

Fig. 7. Resistivity data for transport in the *ab*-plane (a) and along the *c*-axis (b), obtained from the micro-structured crystal shown in Fig. 6.

Fig. 8. Oscillation images about the principal axes of the hexagonal unit cell of the PdRhO$_2$ single crystal chosen for the X-ray diffraction experiment.

Fig. 9. Band structure near the Fermi level $E_F$ = 0 (without spin orbit coupling, rhombohedral Brillouin zone) of PdRhO$_2$ for two different structural data sets, black lines ref. [26], red lines this work.

Fig. 10. (a) Band structure, (b) total and partial densities of states (including spin orbit coupling) of PdRhO$_2$, calculated using structural parameters reported in this paper. The Fermi level is at zero energy.

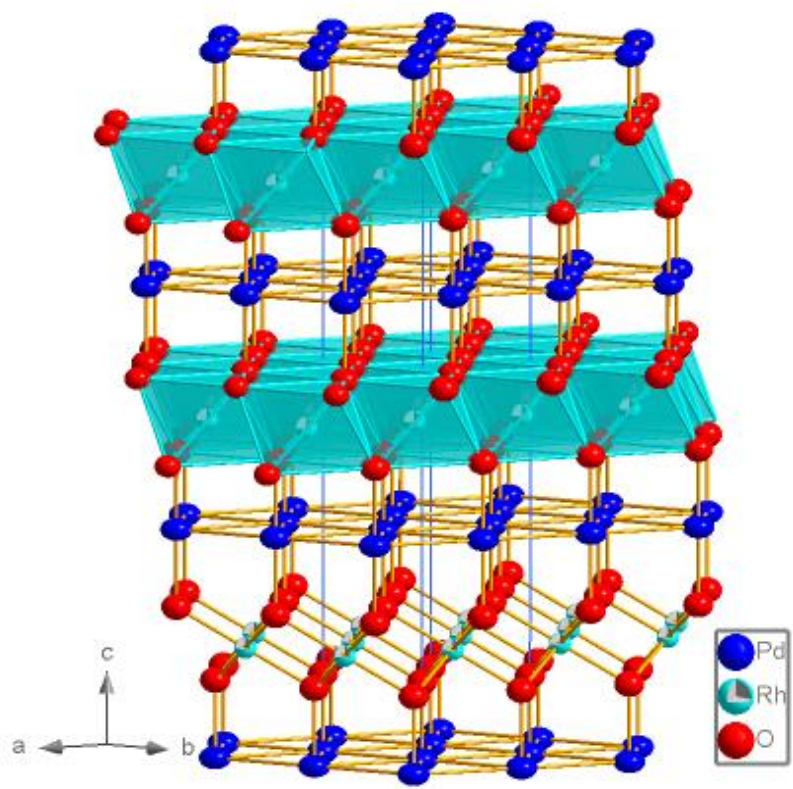

FIG. 1

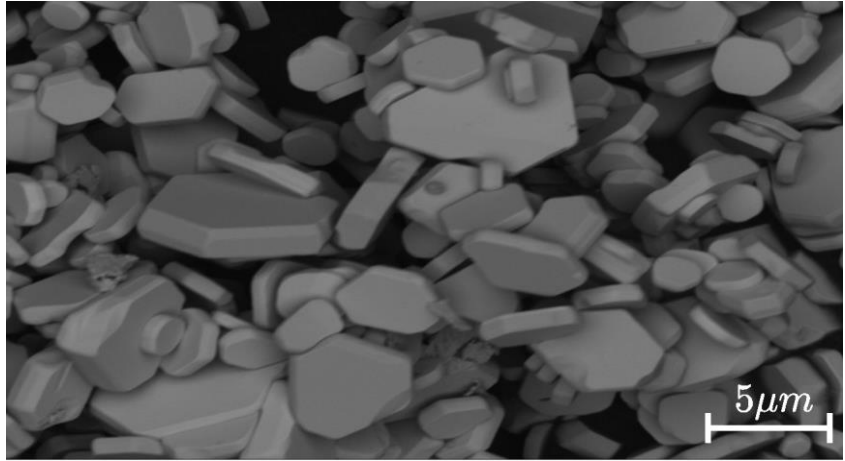

FIG. 2

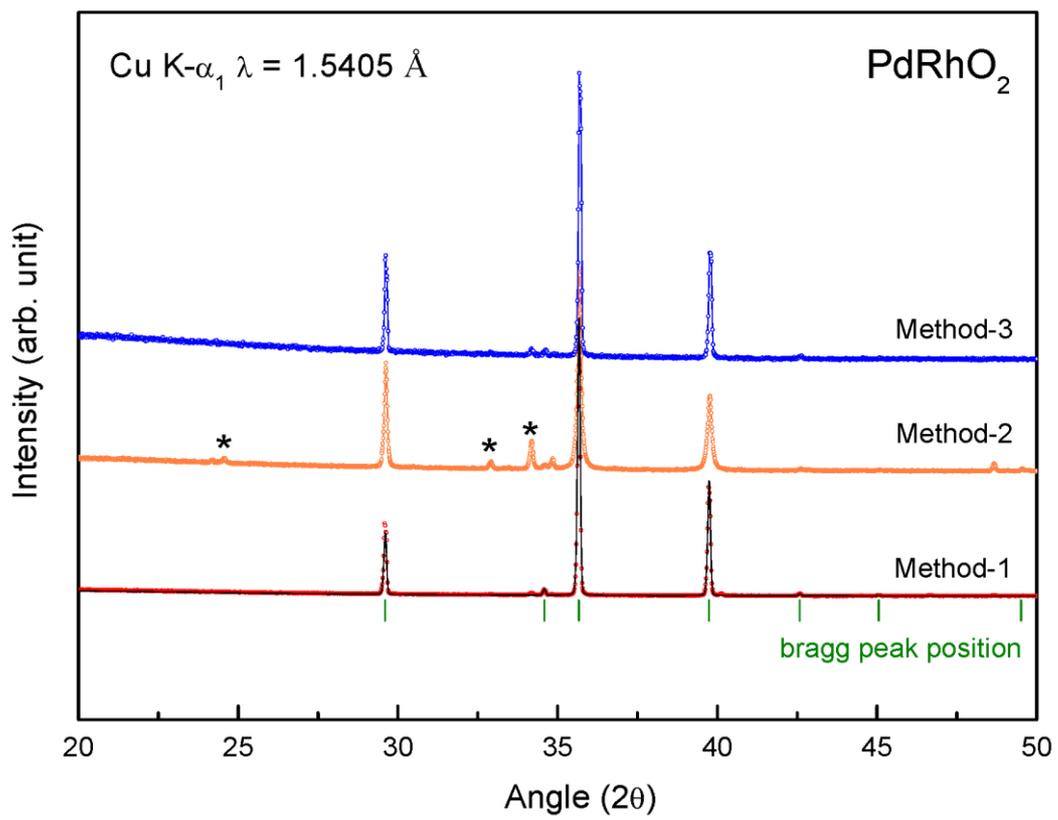

FIG. 3

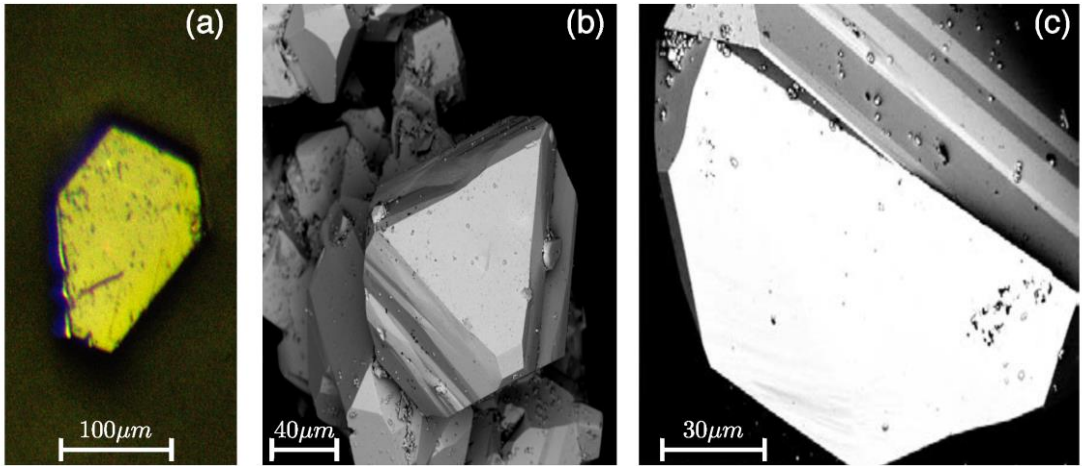

FIG. 4

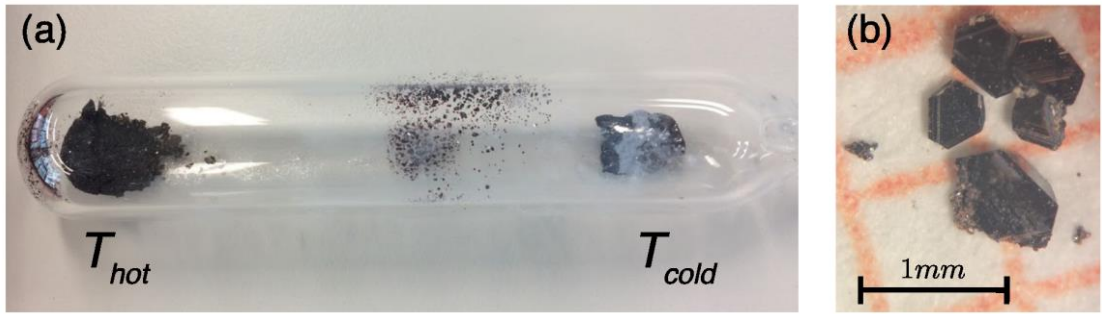

FIG. 5

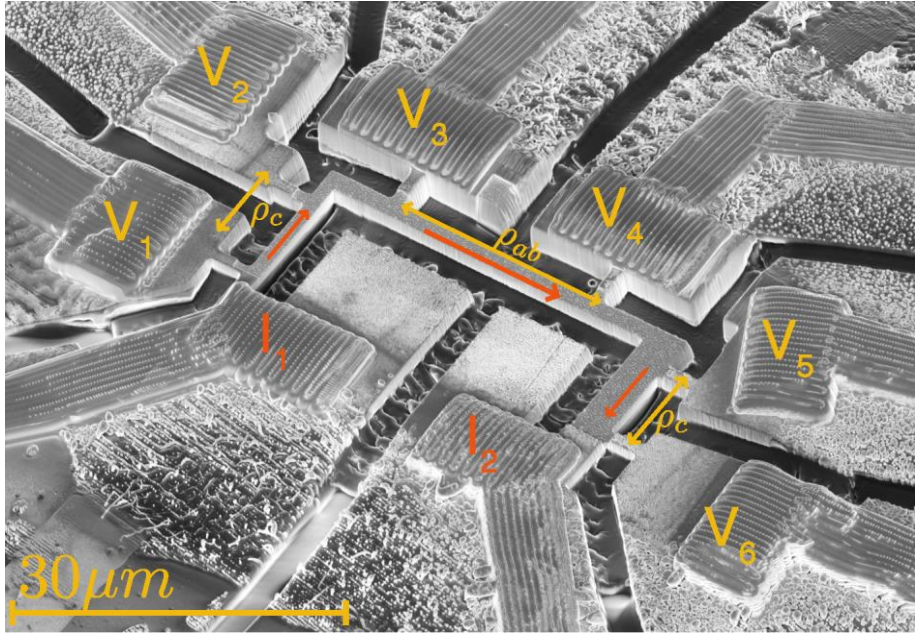

FIG. 6

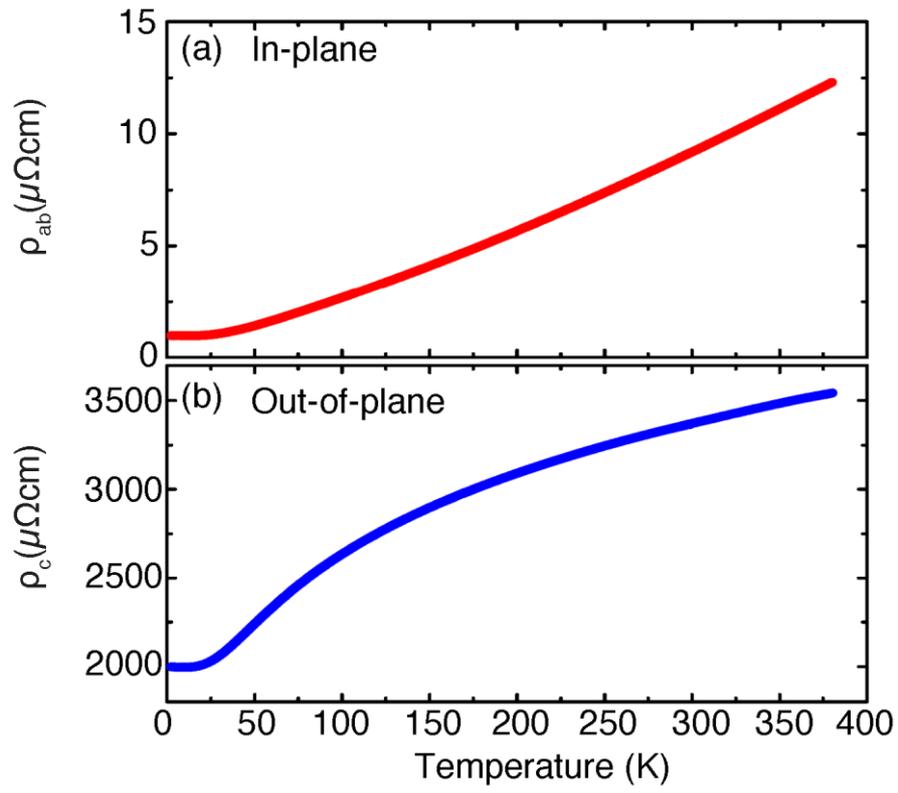

FIG. 7

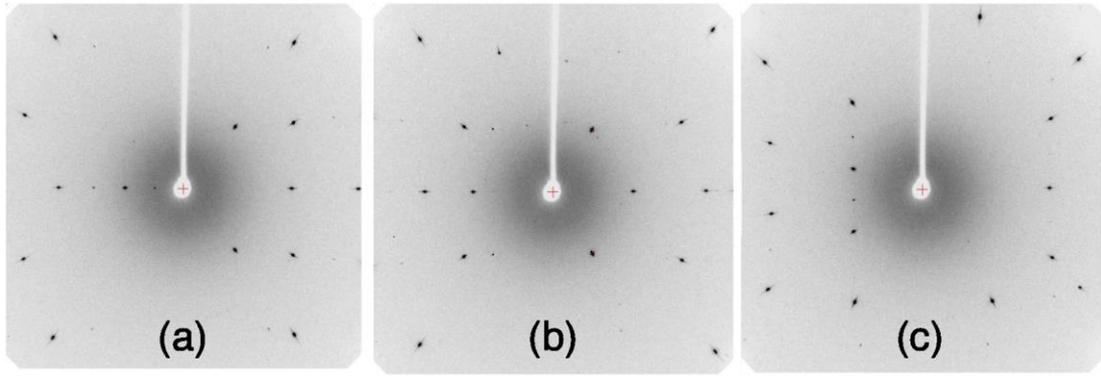

FIG. 8

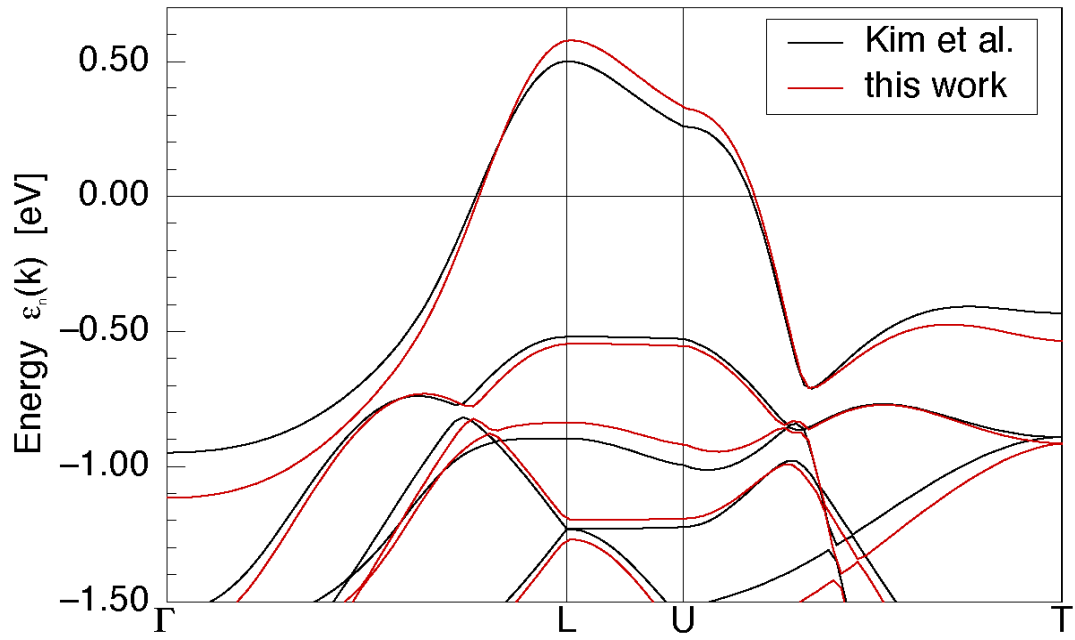

FIG. 9

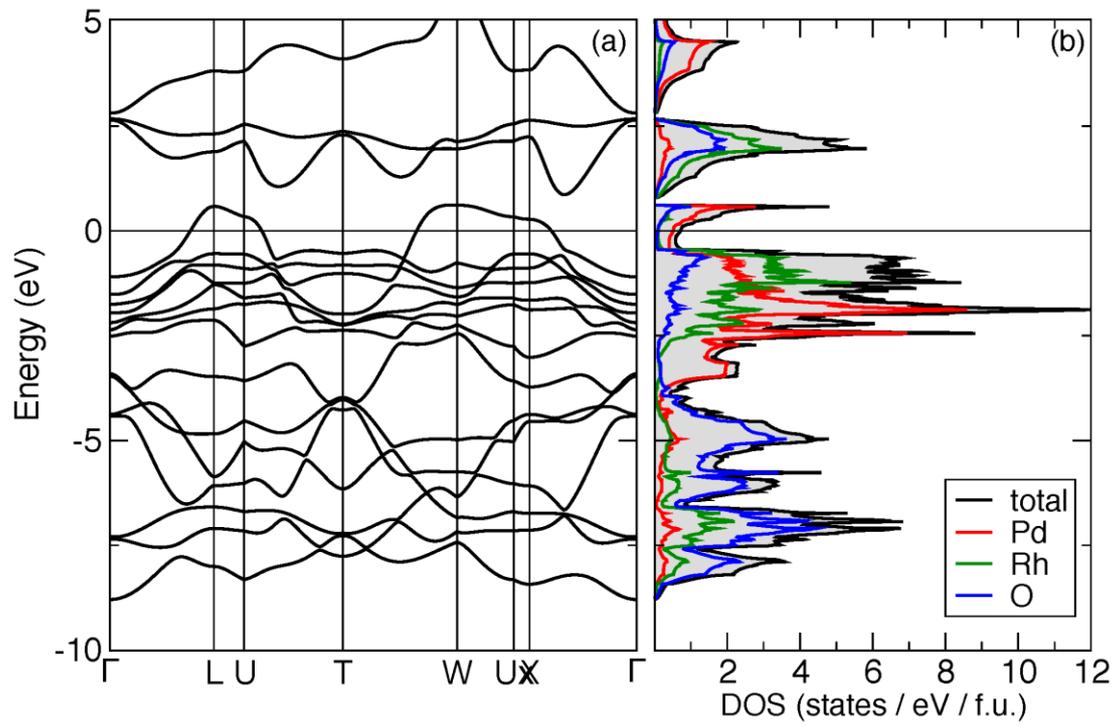

FIG. 10